# Reconstructing Non-Markovian Forces in Atomic Force Microscopy

Sergio Santos

nomos association, Barcelona, Spain

ssantos78h@gmail.com

Abstract

Force reconstruction in AFM leads to quantification of material properties with nanometric resolution but the literature typically relies on testing the different formalisms against well-behaved single-valued functions with no memory kernels. Here, we set to investigate the recovery of forces containing realistic traits such as discontinuities, hysteretic, non-Markovian, i.e., history-dependent tip-trajectories, and non-ergodic, i.e., , trajectory-locking, phenomena. Amplitudes, cantilever stiffness and tip radii have an impact on stability but also on the ability of each method to accurately explore the full tip-trajectory. The small amplitude approximation successfully tracks well behaved and differentiable force components like oscillatory forces of small decay lengths $\lambda < 0.5$ nm even when superposed on highly discontinuous and hysteretic backgrounds. Formalisms relying on Markovian assumptions are shown to recover an "average" force path in the presence of Markovian multivalued functions here modeled as contact hysteresis. For memory dependent forces like capillary interactions, they fully miss the retraction path. In some cases, overlooking assumptions leads to producing material properties based on "artificial" averages while in other cases a real material property is fully missed. Out of the methods explored, only Hooke's law captures complete approach/retraction trajectories and suitably handles discontinuities provided the stability criteria are satisfied.

The Atomic Force Microscope (AFM) was invented exactly four decades ago[1,2]. The general dynamics are monitored as a function of deflection z or tip-sample distance d[3,4] where the unknown tip-sample force controls the behavior. The AFM set-up thus serves as a nanoscale sensor where the target is to monitor, identify and quantify the forces generating contrast to relate them to mechanical[5], chemical[6], electrostatic[7], magnetic[8] or other material properties[9]. The two basic modes of operation are contact and dynamic AFM[4,10]. Force spectroscopy consists in monitoring the "quasi-static" deflection z in contact mode, or the oscillation amplitude, phase or frequency shifts in dynamic modes[4]. Dynamic force reconstruction is complex[2,4,11,12-17]. The results of recovering complex functions however are still to be completely elucidated since most works have used "ideal" [15,18] single-valued and/or differentiable expressions[17,19,20] to validate their results. Here, we set to recover complex and more realistic functions containing discontinuities and hysteretic, non-Markovian, i.e., history-dependent tip-trajectories, and non-ergodic, i.e., , trajectory-locking, phenomena. We explore the parameter space and show how future experimental and theoretical advances should target disentangling and quantifying such mechanisms to extract information via standard set-ups without 1) missing tip trajectories or 2) creating "artificial" averaged material properties.

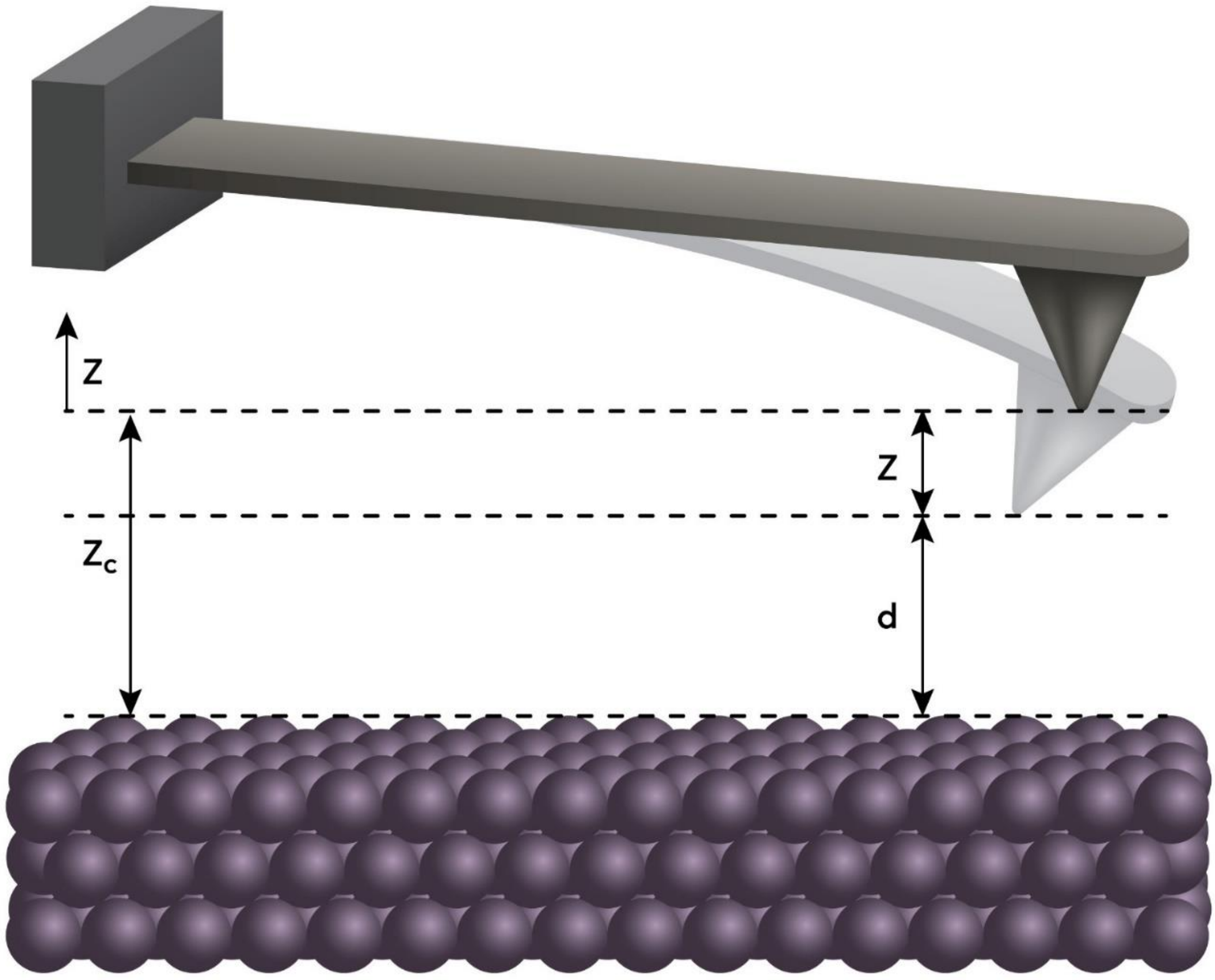

FIG. 1 Illustration of the geometric relationships in a typical AFM set-up.

The equation of motion reduced to a mass m on a spring with constant k and viscosity of the medium $c = \frac{m\omega_0}{Q}$ can be written as

$$m\ddot{z} + \frac{m\omega_0}{Q}\dot{z} + kz = F_0\cos\omega t + F_{ts}[d(t)]_0^\tau \qquad (1)$$

where $F_{ts}$ stands for tip-sample force, z is the cantilever deflection in relation to its position of equilibrium, d is the tip-sample distance, $\omega_0$ is the natural angular frequency, ω is the drive frequency, $F_0$ is the external driving force, and Q is the quality factor. The cantilever-surface separation $z_c$ , z and d are geometrically related (Fig. 1) as follows

$$d \equiv [d(t)]_0^\tau = z_c + z[d(t)]_0^\tau \qquad (2)$$

When the external drive $F_0$>0 (dynamic AFM), in the steady state, the expression can be written as

$$d \equiv [d(t)]_0^\tau = z_c + z_0 + A cos(\omega t + \theta) + \xi([d(t)]_0^\tau) \qquad (3)$$

where $z_0$ is the mean deflection and $\xi[d(t)]_0^\tau$ contains the harmonic distortion that is heretofore disregarded unless otherwise stated. Regarding (1)-(3), terms depending on d are written in terms of a time window τ where

$$[d(t)]_0^\tau \qquad (4)$$

Eq. (4) indicates that, in its most general form, the tip-sample force in AFM must consider a history-dependent functional[21], that is, a functional limit that might be non-Markovian or hereditary, i.e., the system has "memory" and accounts for phenomena such as delayed relaxation, aging, and state changes. Significant examples are capillary interactions[22-24], adhesion hysteresis[25], stick-slip dynamics such as the formation and rupture of weak vdW bonds[26] or elastic stretching and sudden detachment of polymer chains[27], or stiction[28]. If the force is a function of position and velocity[14,15,18,19,29]

$$F_{ts}(z, \dot{z}) = F_c(z) + F_{nc}(z, \dot{z}) \qquad (5)$$

where $\dot{z}$ is the tip velocity and $F_c$ and $F_{nc}$ separate conservative from nonconservative, i.e., dissipative, forces. This ansatz is not a peripheral assumption in many formalisms but foundation upon which the whole frameworks stand on[15,16,18]. We will show that, at best, the assumptions in (5) might recover an “average” force path. At worst, they fully miss a tip trajectory and therefore a targeted material property. Sader et al.[19] proposed the terms even and odd rather than conservative and dissipative or nonconservative respectively but a suitable treatment is still missing[16]. Next we use 1) the small amplitude approximation, 2) frequency modulation (FM), 3) amplitude modulation (AM) and 4) contact mode force reconstruction methods to investigate how fundamental force mechanisms beyond well behaved functions can be recovered.

We start with a continuous, single valued tip-sample force model accounting for vdW interactions via the Hamaker constant H and the tip radius R, and short-range repulsive forces via the DMT model of contact mechanics[4] that takes into account an effective elastic modulus E and the tip-sample indentation δ ($\delta = a_0 - d$)

$$F_{ts}(d) = -\frac{RH}{6d^2} \qquad d > a_0 \qquad (6)$$

$$F_{ts}(d) = -\frac{RH}{6a_0^2} + \frac{4}{3}ER^{1/2}\delta^{3/2} \qquad d \leq a_0 \quad (7)$$

Here, $a_0$ is an intermolecular distance to avoid divergence[4]. The force is shown in Fig 2(a). The interaction in Fig 2(b) is more complex, and it is discussed later.

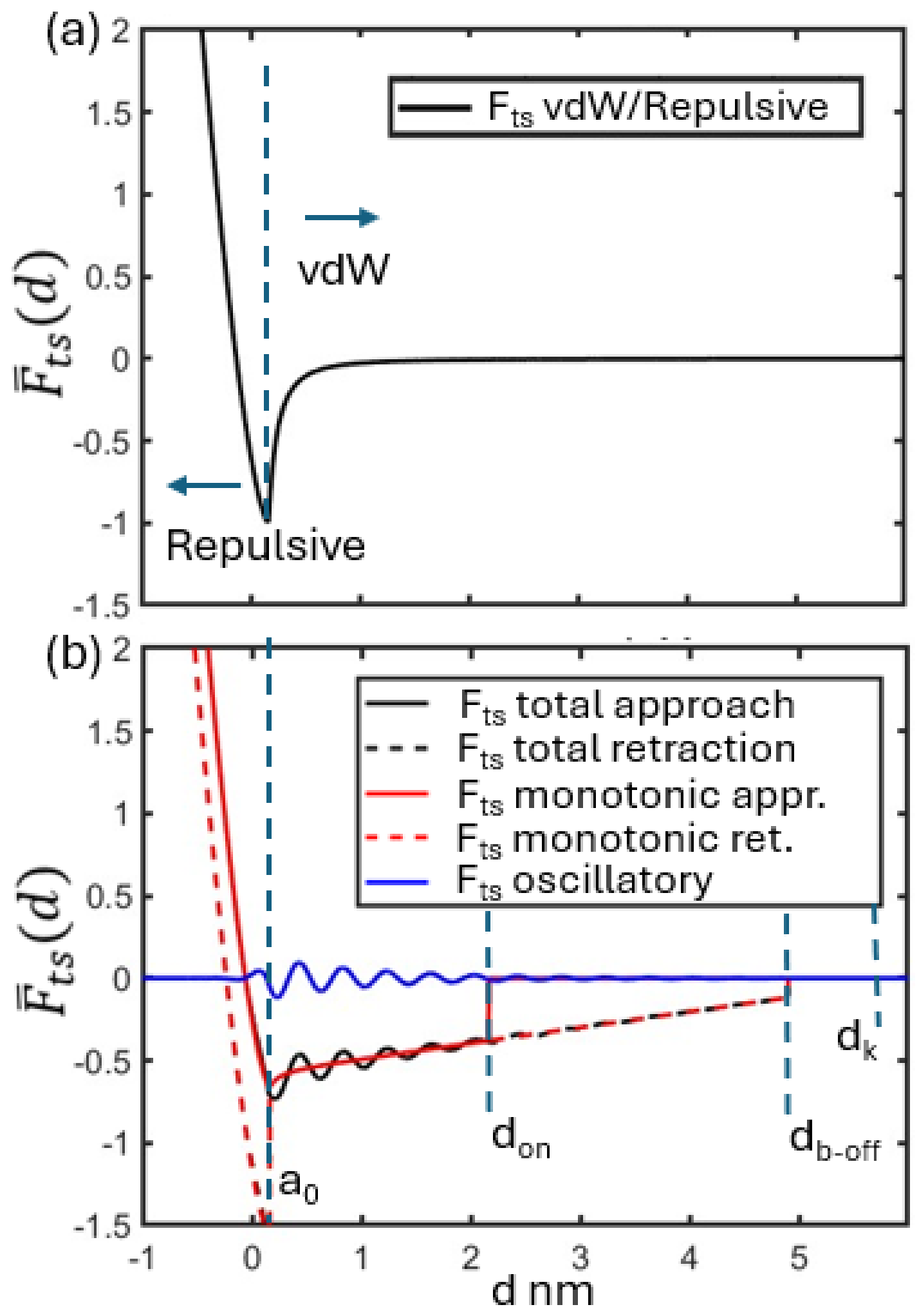

FIG. 2 (a) Single valued force interaction model accounting for vdW and repulsive contact forces. (b) Multi-valued force model that accounts for surface energy hysteresis at contact, capillary forces and an oscillatory force component that can be used to model water structure.

In contact mode AFM the force is reconstructed by monitoring z and using Hooke's law directly - $d = z_c + z$ (Fig. 1). The "instantaneous" force follows from Hooke's law

$$\mathrm{F_{ts}}[d(t)]_0^\tau = -k[z(t)]_0^\tau \qquad (8)$$

In principle Eq (8) is limited to 1) instrumental resolution or signal-to-noise-ratio SNR and 2) the system's stability. Here, all simulations have been carried out "slowly" (~ 10nm/s). The forces employed in the simulations are shown in the figures in black throughout. Fig. 3(a) shows the contact mode results of a simulation where the cantilever-environment-materials parameters are Q=100, $f_0$=70kHz, H=20 zJ, $a_0$=0.165nm and E=1GPa - k and R are shown in the legends. The forces have been normalized everywhere - $\bar{F}_{ts} = \frac{F_{ts}}{|F_{ts}(\min)|}$. Fig. 3(a) shows that for k=4 and R=10 nm information is lost in the attractive and repulsive regimes on approach (blue) and on the attractive on retraction (red). According to Giessibl[17] in contact mode the stability criterion establishes that, for stability at any d

$$k > k_{ts}(d) \qquad (9)$$

where $k_{ts}$ is the stiffness at d. For the current model

$$k_{ts} = \left|\frac{\partial F_{\text{vdw}}}{\partial z}\right| = \frac{HR}{3d^3} > k \qquad \text{snap-in d} \qquad (10)$$

For the values under consideration and the data in Fig. 3(a) the criterion predicts $k \lesssim 15$ N/m for stability. The prediction is that snap-in occurs on approach, $k_{ts} > k$, at d ≈260 pm. Furthermore, the distance for tip-landing on approach can be obtained from

$$k_1\left(d - z_{c,\text{snap-in}}\right) = -\frac{RH}{6a_0^2} + \frac{4}{3}E_{\text{eff}}\sqrt{R}(a_0 - d)^{1.5} \qquad \text{snap-in d on landing} \qquad (11)$$

The solution is $d_{\text{snap-in}} \approx 90\, pm$ (landing). On retraction the first point where Eq. (9) applies coincides with d=$a_0$. This point defines the separation for snap out

$$z_{c,\text{snap-out}} = a_0 - \frac{F_{\text{ts}}(a_0)}{k} = a_0 + \frac{RH}{6ka_0^2} \qquad (12)$$

The same separation defines the landing point on retraction as

$$k_1\left(d - z_{c,\text{snap-out}}\right) = -\frac{RH}{6d^2} \qquad \text{snap-out d on landing (13)}$$

In summary from Eqs. (11) and (12) information is lost on approach for 90 pm <d <260 and on retraction for $a_0$ <d< 420pm. The snap in $z_{c,\text{snap-in}}$ and snap out separations $z_{c,\text{snap-out}}$ are also shown in Fig. 3(b). The predictions and the figure illustrate that the stability criterion is deterministic and can be found for analytical models. The analytical results are also in good agreement with the simulations in Figs. 3(a) and (b)[16]. There are at least two ways in which cantilever-tip parameters can be selected to overcome the stability problem. The trivial one follows directly from Eq. (9) and involves increasing k. The second is to reduce R since it controls the magnitude of attractive forces. According to the criterion the minimum k for stability for R= 10nm is approximately 15 N/m, while for a k=4N/m, R ≈ 3nm. The simulations in Fig. 3(c) show that stability follows for k=20 Nm and R=10 nm (blue) and for k=4N/m and R=2.5nm (red).

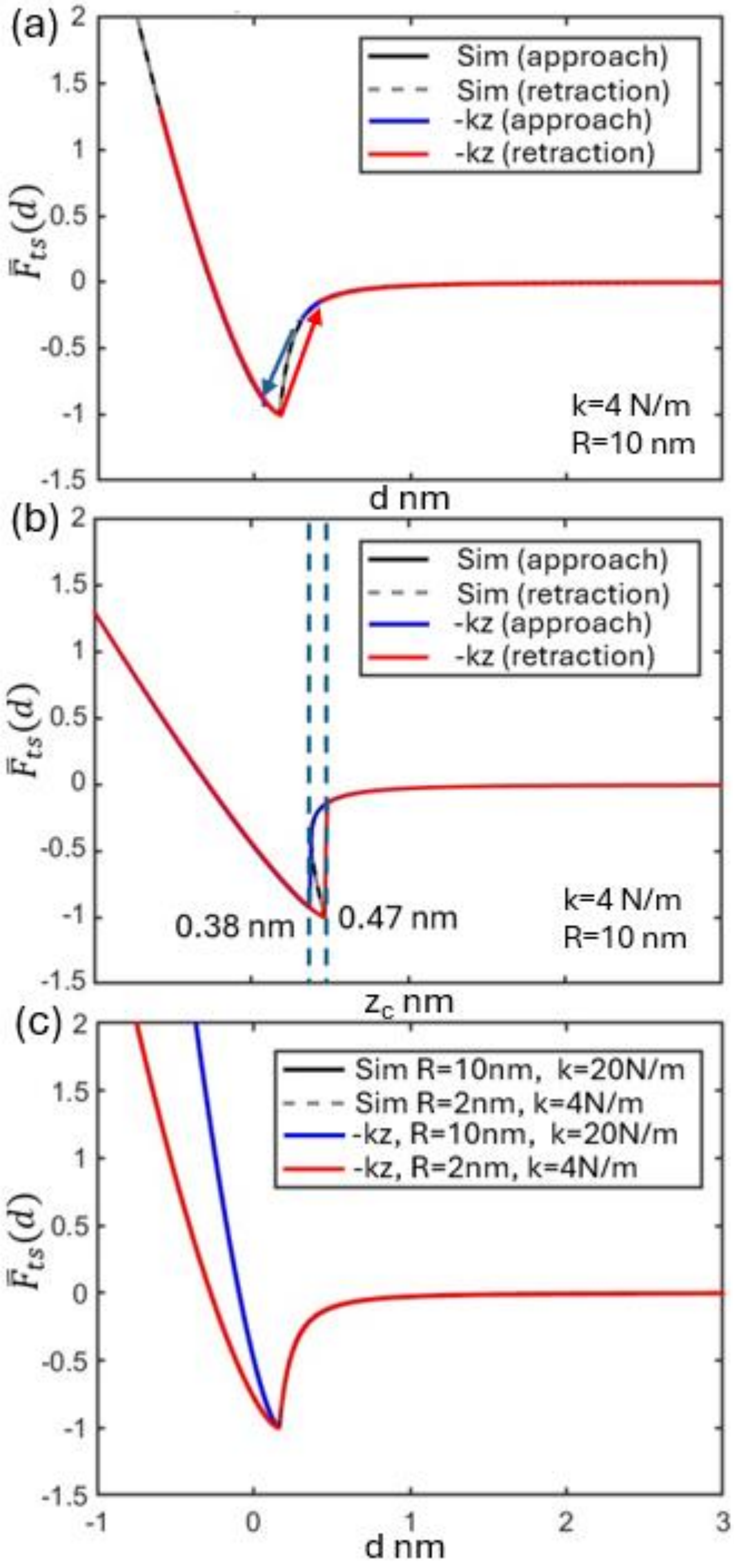

FIG. 3. Contact mode simulations for the recovery of the force displaying system instabilities in terms of (a) d and (b) $z_c$. (c) Stability can be recovered by increasing k (blue lines) or decreasing R (red lines).

Simulation results are shown in Fig. 4 in FM AFM (small amplitude approximation) for the same model (Fig. 2(a)). Linearizing at a distance $d_0 = z_c + z$

$$m\ddot{z} + \frac{m\omega_0}{Q}\dot{z} + (k + k_{ts})z = F_0(\mathrm{d})\cos(\omega t) + F_{ts}(d_0) \qquad (14)$$

where the term $F_{ts}(d_0)$ introduces a constant offset, leading to

$$\frac{\Delta f}{f_0} \equiv \Omega = \sqrt{\frac{k+k_{ts}(d)}{k} - \frac{1}{4Q^2}} - 1 \qquad (15)$$

Provided A is sufficiently small

$$k_{ts}(d_0) \approx \frac{1}{2k}\frac{\Delta f}{f_0} = \frac{1}{2k}\Omega \qquad (16)$$

The above are well-known results. In FM the frequency is experimentally monitored so Eq. (16) can be solved to reconstruct $F_{ts}$. When the cantilever is sinusoidally driven[17,30] the extended stability criterion gives

$$|\ kA_0(F_{ts}(\mathrm{d}))\ | > |\ \text{atractive force}\ | \quad (17)$$

Results are shown in Fig. 4(a) for the same parameters used in Fig. 3(a) and k=4N/m, $A_0$= 50pm and R=10 nm. Eq. (17) establishes that the system is unstable since 25 N/m $\lesssim$ k for stability. This agrees with the results in Fig 4(a) where the force has been recovered from (16) (blue lines) and via Hooke's law (red lines). From (17) $A_0$ becomes a significant parameter beyond k and R. While increasing $A_0$ leads to stability, a limitation here is the small amplitude approximation since it is required that $A_0 \ll \lambda$ (decay length of the force). In the model (Fig. 2(a)) $\lambda \sim 0.1$-1. The tension arises from the requirement that $A_0 \lesssim 0.1$ for linearity and $kA_0(F_{ts}(\mathrm{d})) > \text{atractive force}$ for stability. From Eq. (17) and for k=4N/m and R=10nm (other parameters as in Fig. 2(a)) the extended criterion predicts $A_0 > 300$pm. Simulations show (not shown) that $A_0$=100pm still leads to instabilities while for $A_0$=400pm linearization breaks down. Again, two possibilities follow. First decreasing R. Second, increasing k. From Eq. (17) and k=4 N/m, R$\lesssim$ 2.7 nm for stability. For R=10 nm, 25 N/m $\lesssim$ k for stability. These values indicates that FM might require larger values of k than contact mode for stability since, from before, 15

N/m ≲ k (stability criterion) for contact mode and 25 N/m ≲ k (extended stability criterion) for $A_0$=50pm. Fig. 4(b) shows that stability is reached at k=30N/m for $A_0$=50pm and R=10nm. It is worth noting, however, that small values of k also lead to errors in the repulsive regime. This is indicated in Fig. 4(a) with an arrow. Linearization can be extended to AM AFM (see Eq. (18)) by recovering the resonant frequency from the observables A and θ. On the other hand, in AM AFM, A varies and typically rapidly decreases with decreasing separation. For the small amplitudes, i.e., $A_0$=50-100pm (not shown), required for linearization and for the type of forces under consideration, the amplitude decays to a few pm much before the tip reaches repulsive forces, i.e., the system falls outside the detection range. A solution is to increase k, i.e., ~1k (not shown).

.

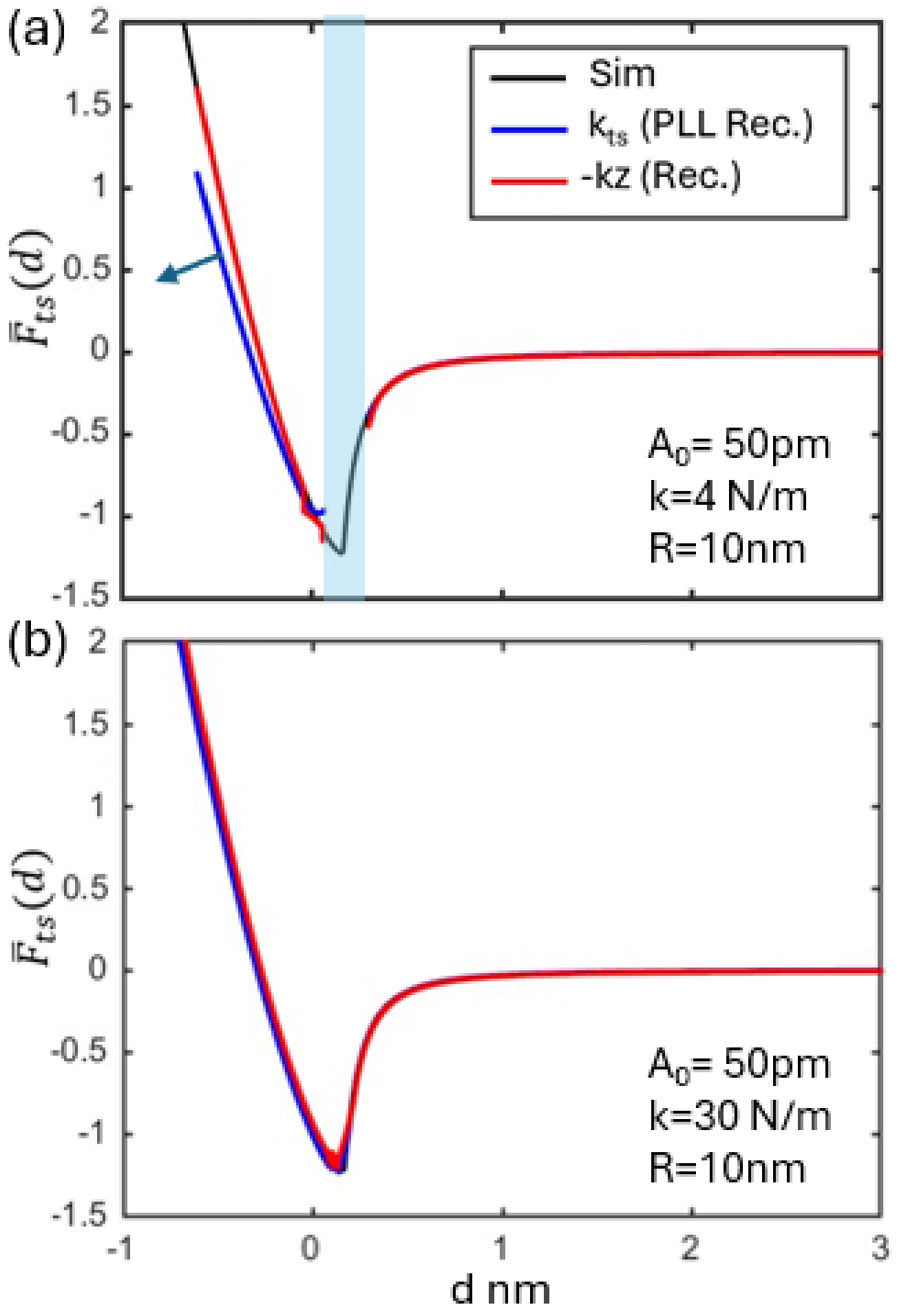

FIG. 4 (a) FM AFM simulation for a linearized system displaying instabilities. (b) k is increased beyond the valued predicted the extended stability criterion. Both the linearized method of reconstruction (blue lines) and Hooke's law (red lines) are shown.

The Sader-Jarvis (SJ) formalism[12] was extended to AM by Katan[13] – more recently other similar formalisms have been developed[14] – where $A_0$ is the free amplitude, A and θ vary with d and the normalized resonant frequency Ω can be obtained as [13]

$$\Omega(d) = \left[1 + \frac{A_0}{QA(d)} cos(\theta(d))\right]^{1/2} - 1 \qquad (18)$$

The force is[19]

$$\bar{F}_{ts}(d) = \frac{2k}{|\,F_{ts}(min)|} \int_{u=d}^{u=\infty} \left[\left(1 + \frac{A^{1/2}}{8\sqrt{\pi(u-d)}}\right)\Omega(u) - \frac{A^{3/2}}{\sqrt{2(u-d)}}\frac{d\Omega(u)}{du}\right] du \qquad (19)$$

Results are shown in Fig. 5(a) for $A_0$=40nm (red lines) and $A_0$=10nm (blue lines). Instabilities are observed for $A_0$=10 nm in AM AFM even when the extended stability criterion predicts that 300 pm $\lesssim A_0$ for stability (see discussion of Fig. 6 below). This follows from a main difference between AM and FM where A decays with the force in AM. Our simulations indicate (not shown) that $A_0$ > 25-30 nm for these conditions (R=10nm and k=4N/m) for stability. We further find that there is an almost linear relationship between increasing k or $A_0$ and decreasing R to reach stability. This is shown in Fig. 5(b) where $A_0$ was divided by 4 and either R was divided by 4 (blue lines) or k was multiplied by 4 (red lines) for stability and full recovery of that range.

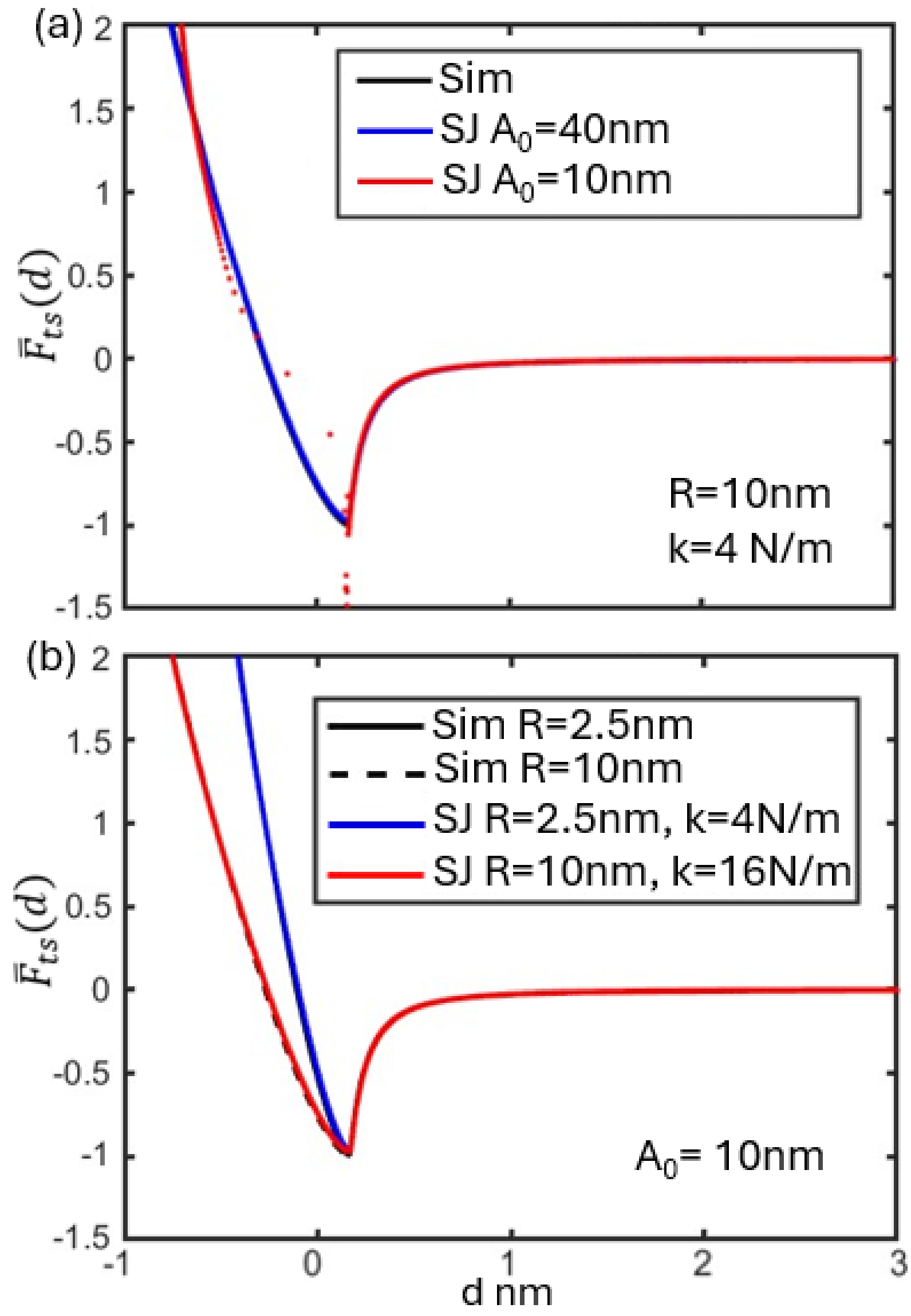

FIG 5. Force reconstruction in AM AFM (Sader-Jarvis Eq. (19)) showing (a) instabilities for smaller values of $A_0$ (red lines) and stability for larger values (blue lines). (b) Stability follows for the same system by either decreasing R (blue lines) or increasing k (red lines).

In FM AFM Ω is recovered directly and A is constant, i.e., $A=A_0$. Otherwise (19) applies directly. For the same values of k and R as those in Fig. 5(a), stability is achieved (Fig. 6) for values of $A_0$ as small as 2nm (and lower – not shown) in FM AFM (25-30 nm in AM AFM). In short, while the extended stability criterion is a simplification of the phenomenon its predictions are reasonable for FM AFM.

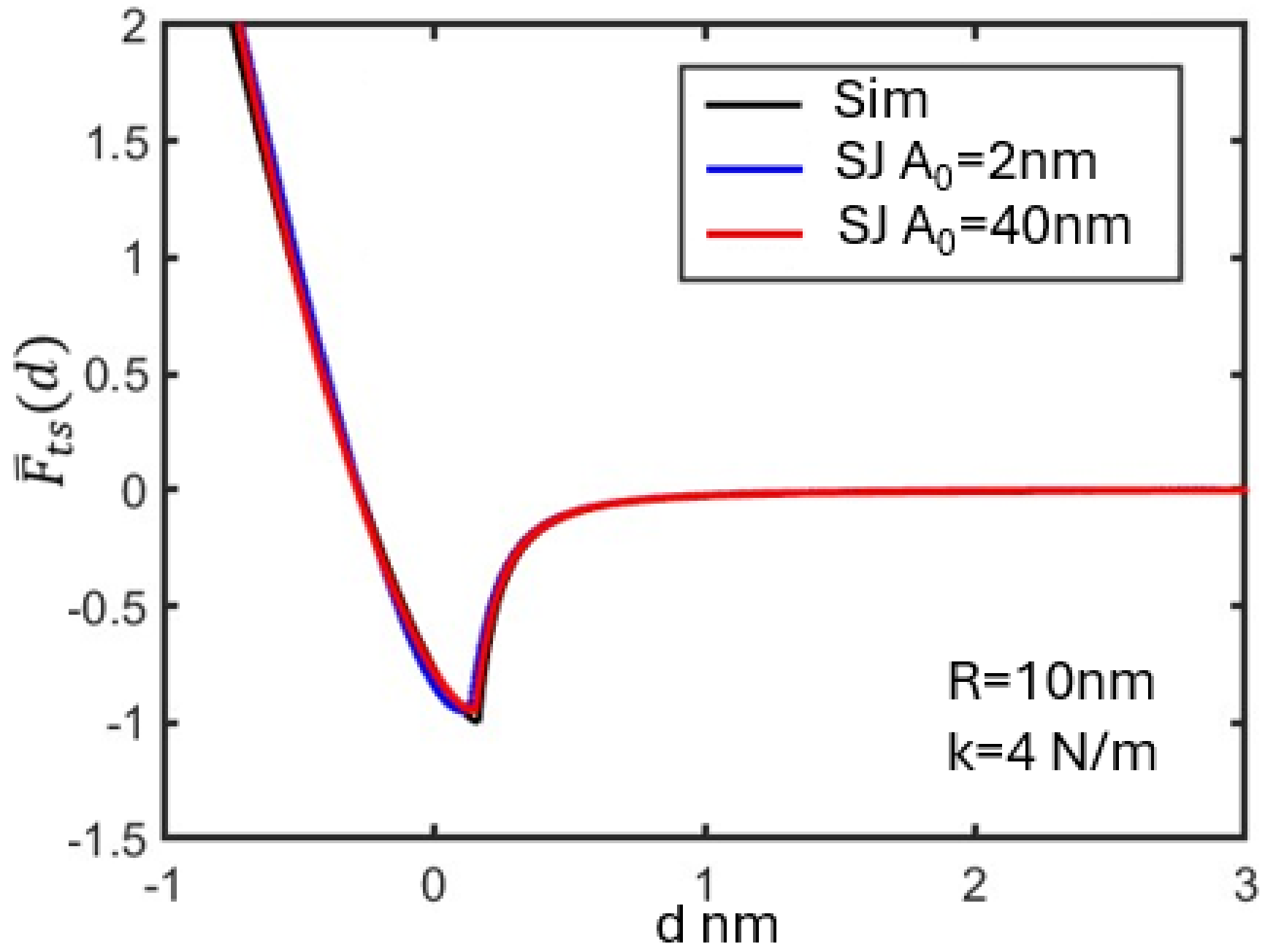

FIG 6. Force reconstruction in FM AFM (Sader-Jarvis Eq. (19)) showing that the system is already stable for values of $A_0$ as low as $A_0$= 2nm (blue lines) for the same parameters as those used in AM AFM in Fig. 5(a), even if it slightly improves for larger values (red lines).

We now move to the complex forces shown in Fig. 2(b). First, in the contact region, hysteresis can be modeled as

$$F_{ts}(d) = -\frac{R}{6a_0^2}(H + \Delta H) \equiv -4\pi R(\gamma_a + \Delta\gamma)\ , \ d \leq a_0 \qquad (20)$$

where $\gamma_a$ is the surface energy of the tip-sample contact on approach. Surface energy increases by $\Delta\gamma/\Delta H$ on retraction leading to irreversible energy loss[31]. The approach (continuous red lines) and retraction (dashed red lines) paths, and the force discontinuities, in Eq. (20) are shown in Fig. 2(b) for $d<a_0$. In ambient conditions a capillary force may also be present. The Kelvin equation dictates[22,32] that, in the constant chemical approximation the capillary forms at a distance $d_{on}$ ($d_{on}\approx 2h$ where h is the height of a water layer[32]) and breaks at a distance $d_{off}$ ($d_{off}> d_{on}$)[22,32]

$$F_{ts}(d) = -4\pi R\gamma_w \cos(\vartheta)(1-\frac{d}{d_K}) \quad \mathrm{d} > a_0, \quad \mathrm{d}=a_0 \text{ for } \mathrm{d} \leq a_0 \quad (21)$$

Here $\gamma_w$ is the surface energy of water, $\vartheta$ is the contact angle and $d_k$ is the kelvin distance where the capillary bridge "ideally" breaks ($d_{off} \equiv d_k$)[33]. Fig. 2(b) shows a capillary discontinuity at $d_{on}$ and a second discontinuity on retraction where $d_{off} \neq d_k$ - $d_{off} \equiv d_{b\text{-}off} < d_k$. Finally, inside the capillary, water structure may lead to oscillatory forces[24,34]

$$F_{osc}(d) = \begin{cases} A_{osc} cos\left(\frac{2\pi(\mathrm{d}-a_0)}{\sigma_{osc}} + \varphi_{osc}\right) \cdot e^{-\frac{\mathrm{d}}{\lambda_{osc}}} & for\ \mathrm{d} \geq a_0 \\ A_{osc} cos\left(\frac{2\pi(\mathrm{d}-a_0)}{\sigma_{osc}} + \varphi_{osc}\right) \cdot e^{-\left(\frac{\mathrm{d}}{\lambda_{osc}} + \frac{a_0-\mathrm{d}}{\lambda_{fade}}\right)} & for\ \mathrm{d} < a_0 \end{cases} \quad (22)$$

where osc stands for oscillatory. The force is shown (blue lines) in Fig. 2(b). The oscillatory force is continuous and only applies when the capillary force (21) is active. Forces (6), (7) and (20)-(22) have been used in the simulations leading to Figs. (7)-(9) - H=0.1zJ, k=4N/m, $\Delta\gamma = 10\gamma_a$, h=1nm, $d_{b\text{-}off} = 0.8d_k$, $\vartheta$=85, $A_{osc}$ =0.1nm, $\varphi_{osc}$=90, and $\lambda_{fade}$=0.1nm (rest as in Figs. 3-6 or as described for each simulation). In summary, Eq. (20) is both Markovian and ergodic and includes discontinuities on approach and retraction, the capillary force from Eq. (21) is non-Markovian and non-ergodic and includes discontinuities and Eq. (22) includes forces with short decay lengths, i.e., $\lambda_{osc}$<0.5-1nm.

In Fig. 7 the force in Fig 2(b) is reconstructed using Hooke's law (method from Eq. (8)) for $\sigma_{osc}$=0.4nm, $\lambda_{osc}$=1nm and $d_{b\text{-}off} = d_k$ (for Figs. 8 and 9 $\sigma_{osc}$=0.5nm and $\lambda_{osc}$=2nm). Inspection shows that this method successfully recovers all forces by inducing only minimal errors at the discontinuities. Provided the stability criterion is satisfied (Eq. (9)) and there is sufficient SNR, the whole set of phenomena can be recovered in contact mode from Hooke's law. For Fig. 7 $k_{ts}(maximum) \approx 1.2$ < k=4 N/m agreeing with the observed stability.

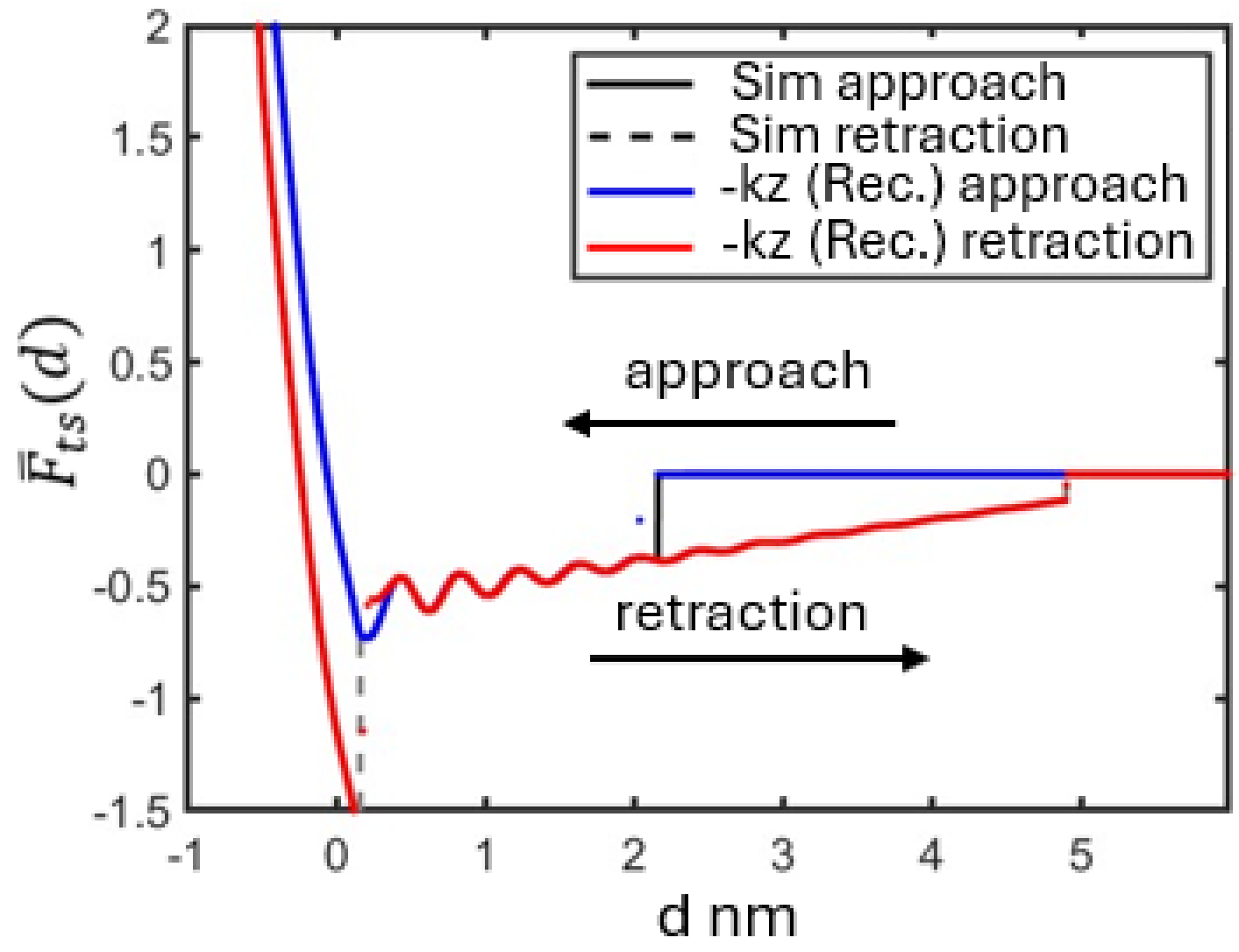

FIG 7. Contact mode (Hooke's law) successfully recovers all phenomena in Fig. 2(b) on tip approach (blue lines) and retraction (red lines).

The results of the small amplitude approximation for FM AFM (Eqs. (14)-(16)) are shown (red lines) in Fig. 8 on (a) approach and (b) retraction for $A_0$= 50pm for the forces in Fig. 2(b) – Hooke's law is also shown (blue lines). Some problems can be identified. First, linearity implies continuity since the derivation requires $\frac{\delta F_{ts}}{\delta d}$ to exist. Second, the extended stability criterion in Eq. (17) cannot account for the discontinuities. Third, even if it did, the attractive force exceeds 1.5nN implying the minimum $A_0$ for k= 4N/m should be ≈ 400pm ~ $\sigma_{osc}$ (loss of linearity). Again, k could be increased or R reduced, with the disadvantages already discussed for the results in Fig. 4 and the vdW model. Still the small amplitude method resolves (Fig. 8) the oscillation peaks ≈ 500pm= $\sigma_{osc}$ (red lines) on both approach and retraction and only significantly worsens the contact mode signal (blue lines) around the point of contact $a_0$.

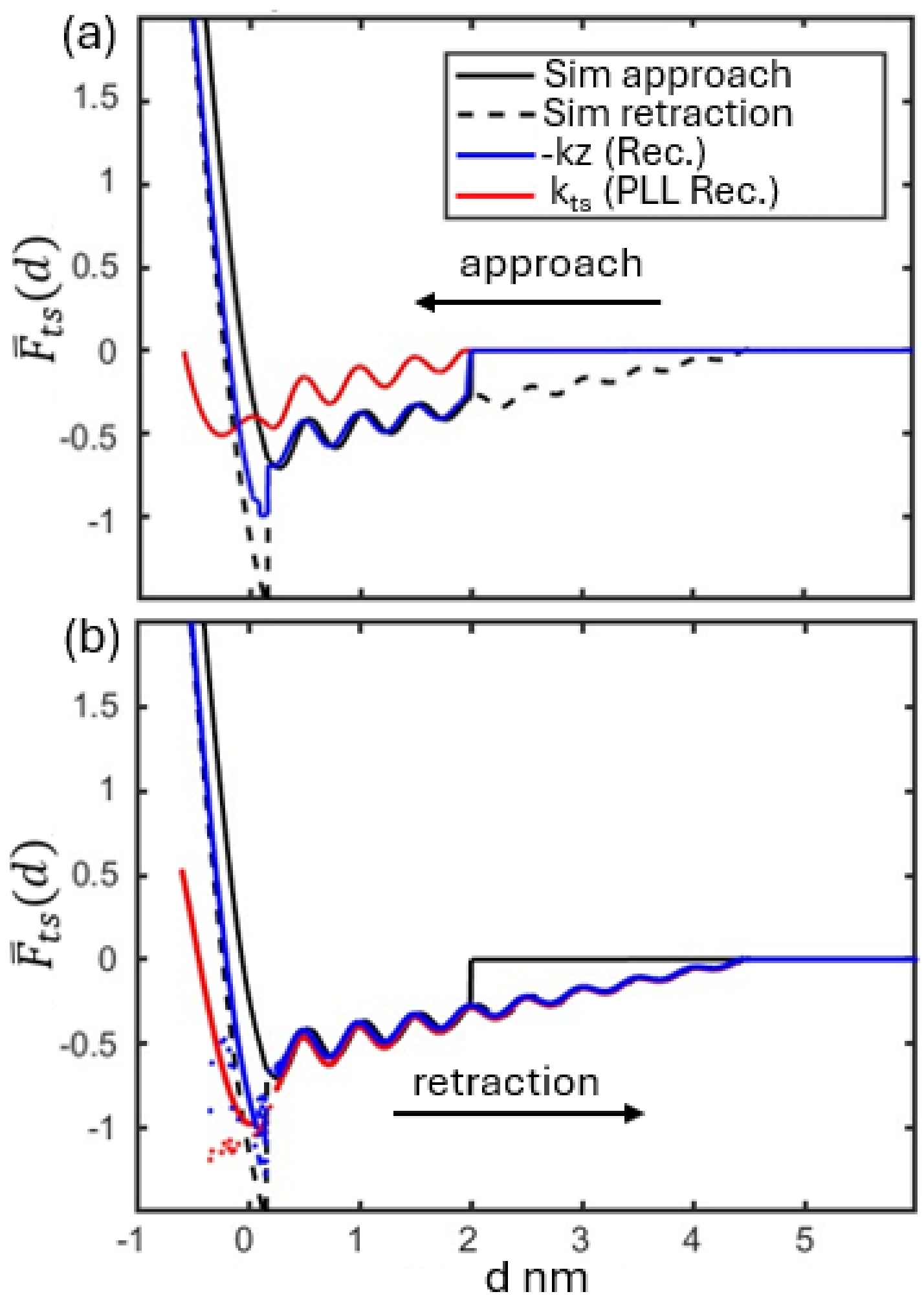

FIG 8. FM AFM simulation for a linearized system for the forces modelled in Fig. 2(b) (the same as those in Fig. 7). (a) and (b) show the results of tip approach and retraction for the linearized reconstruction method (red lines) and Hooke's law (blue lines), via the mean deflection signal.

The results of simulations for the Sader-Jarvis (SJ) formalism are shown in Fig. 9 for FM AFM (blue lines) and AM AFM (red lines). First, the data shows that multi-valued functions such as contact hysteresis - Eq. (20) - produce an average. This has implications when the target is monitoring the difference between the approach and retraction paths.

Second, for non-Markovian mechanisms such as the capillary force only the approach path is recovered. This is independent of whether the approach or retraction data is employed (data not shown). Third, again, FM AFM can lead to stable results for lower values of $A_0$ than AM AFM. For the data in Fig 9 $A_0$ was stable at $A_0 \approx 0.5$nm (not shown) where the extended stability criterion predicted approximately 0.3 nm. In AM AFM $A_0 \approx$ 25-30 nm (not all data shown). The oscillatory behavior on the other hand, is recovered by both FM and AM. Finally, the Gibbs phenomenon is observed at the points of discontinuity. This is a problem of any method relying on differentiability at a point. On the other hand, unlike the small amplitude approx. (Fig. 8), these formalisms manage to recover the absolute value of the forces after recovering from the Gibbs phenomenon.

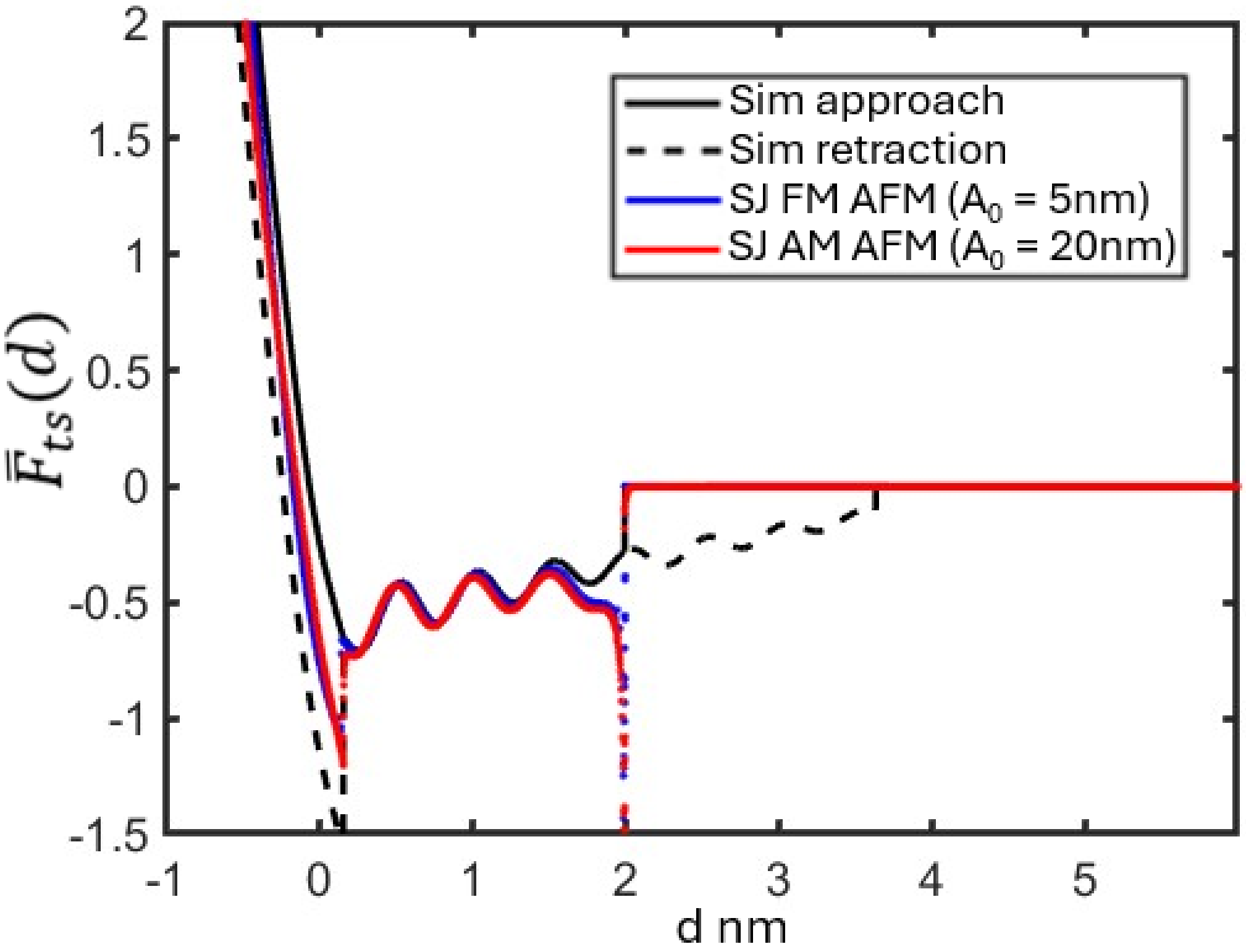

FIG 9. Force reconstruction (Sader-Jarvis Eq. (19)) in FM AFM (blue lines) and AM AFM (red lines) showing the behavior of the formalisms when dealing with discontinuities and hysteretic mechanisms involving Markovian (contact hysteresis Eq. (20)) and non-Markovian (capillary hysteresis Eq. (21)) behavior.

We have submitted Hooke's law, the small amplitude approximation and the Sader-Jarvis reconstruction formalisms in the AM and FM modes to complexities beyond single-valued, Markovian and velocity dependent forces and tested stability criteria. While relying on the "even-odd" mathematical assumptions has led to the robust derivation of force reconstruction methods in dynamic AFM, users must be aware that the reconstructed forces might include the creation of averaged tip-trajectories resulting in the "artificial" creation material properties or, in the worst case, they might miss the targeted properties. Out of the methods explored, only Hooke's law captures complete approach/retraction trajectories and suitably handles discontinuities provided the stability criteria are satisfied.

## Data Availability Statement

The data that support the findings of this article are available upon request.

## Acknowledgments

Maritsa Kissamitaki produced the artistic illustrations.